\documentclass[11pt, a4paper]{article}

\usepackage{fullpage}

\usepackage{epsfig}
\usepackage{amssymb}
\usepackage{amsmath}
\usepackage{multirow}

\usepackage{graphicx}

\usepackage{caption}
\usepackage{subcaption}
\usepackage{filecontents}

\begin{document}

\title{Feature Extraction from Degree Distribution for Comparison and Analysis of Complex Networks}

\author{Sadegh Aliakbary, Jafar Habibi, and Ali Movaghar\\
Department of Computer Engineering, Sharif University of Technology, Tehran, Iran\\
aliakbary@ce.sharif.edu, jhabibi@sharif.edu, movaghar@sharif.edu}

\maketitle



\begin{abstract}
The degree distribution is an important characteristic of complex networks. In many data analysis applications, the networks should be represented as fixed-length feature vectors and therefore the feature extraction from the degree distribution is a necessary step. Moreover, many applications need a similarity function for comparison of complex networks based on their degree distributions. Such a similarity measure has many applications including classification and clustering of network instances, evaluation of network sampling methods, anomaly detection, and study of epidemic dynamics. The existing methods are unable to effectively capture the similarity of degree distributions, particularly when the corresponding networks have different sizes. Based on our observations about the structure of the degree distributions in networks over time, we propose a feature extraction and a similarity function for the degree distributions in complex networks. We propose to calculate the feature values based on the mean and standard deviation of the node degrees in order to decrease the effect of the network size on the extracted features. The proposed method is evaluated using different artificial and real network datasets, and it outperforms the state of the art methods with respect to the accuracy of the distance function and the effectiveness of the extracted features.

\end{abstract}

\maketitle 

\section{Introduction}
\label{Introduction}
Degree distribution is an important and informative characteristic of a complex network \cite{ref1,ref2,ref3,ref4,ref7,ref8,ref9,ref10,ref11,ref13,ref16,ref19,ref22, ref27,ref28,ref29,ref30}. Although the degree distribution does not capture all aspects of the topology of a network \cite{ref18}, it reflects the overall pattern of connections \cite{ref27} and is an important determinant of network properties \cite{ref28}. Degree distribution is also a sign of link formation process in the network. The degree distribution of complex networks often follows a heavy-tailed distribution \cite{ref8,ref9,ref10,ref11,ref12}, but networks still show different characteristics in their degree distributions. Hence, we frequently need an appropriate similarity function in order to compare the degree distribution of the complex networks. Such a similarity function plays an important role in many network analysis applications, such as evaluation of network generation models \cite{ref1,ref2,ref5,ref7,ref9,ref29}, evaluation of sampling methods \cite{ref11,ref15,ref16,ref17,ref18}, generative model selection \cite{ref61,ref20,ref21}, classification or clustering of network instances \cite{ref3,ref19,ref20,ref21,ref61}, anomaly detection \cite{ref32,ref33}, and study of epidemic dynamics \cite{ref34,ref35,ref36}. For example, in evaluation of sampling algorithms, the given network instance is compared with its sampled counterpart in order to ensure that the structure of the degree distribution is preserved \cite{ref11,ref15,ref16,ref17,ref18,ref19}. 

In addition to the need for network comparison, representing the network as a fixed-size feature vector is also an important step in every data analysis process \cite{ref61,ref20,ref21}. In order to employ the degree distribution in such applications, a procedure is needed for extracting a feature vector from the degree distribution. The extracted feature vector is also useful in developing a distance function for comparing two degree distributions. Although there exist accepted quantification methods for many network features (e.g., average clustering coefficient, modularity, and average shortest path length), the quantification and comparison of the degree distribution is not a trivial task. We will show that the existing methods have major weaknesses in feature extraction and comparison of networks, particularly when the considered networks have different sizes. Even if no network comparison is required, feature extraction from the degree distributions has independent applications. For example, data-analysis algorithms require the network features, including the degree distribution, to be represented as some real numbers in the form of a fixed-length feature vector \cite{ref61,ref19,ref20,ref21}.

According to the mentioned applications for the demanded similarity function, we regard two degree distributions similar if their networks follow similar structure, similar connection patterns and similar link formation processes, even if the networks are of completely different scales and sizes. The state of the art approaches for comparing degree distributions are eye-balling the distribution diagrams (usually, to satisfy a heavy-tailed distribution)  \cite{ref2,ref9,ref22}, Kolmogorov-Smirnov (KS) test  \cite{ref8,ref10,ref11,ref23,ref24,ref25}, comparison based on fitted power-law exponent \cite{ref22,ref26}, and comparison based on distribution percentiles \cite{ref20}. Eyeballing is obviously an inaccurate, error-prone and manual task. Comparison based on power-law exponent is based on the assumption that the degree distributions obeys a power-law, which is invalid for many complex networks \cite{ref13,ref14,ref38,ref39}. KS-test is based on a point-to-point comparison of the cumulative distribution function, which is not a good approach for comparing networks with different ranges of node degrees. Percentile method is too sensitive to the outlier values of node degrees. As a result, the existing methods are actually inappropriate for comparing the degree distribution of networks.

\begin{figure}
	\centering
	\includegraphics[scale=0.45]{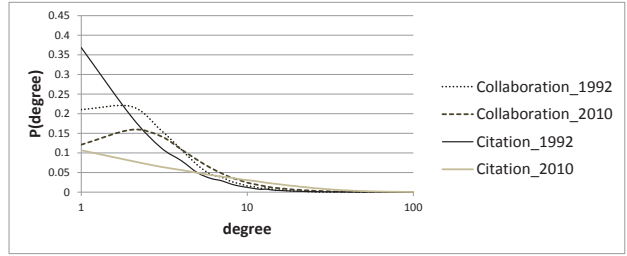}
	\caption{ \label{fig:cit_col} Degree distribution of two citation networks and two collaboration networks}
\end{figure}

\begin{figure}
	\centering
	\includegraphics[scale=0.45]{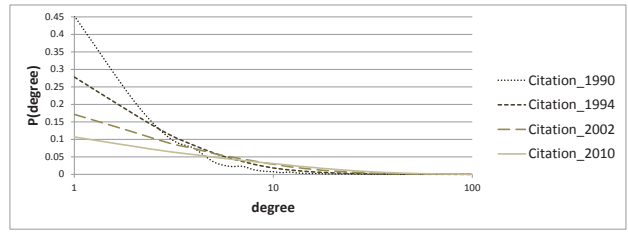}
	\caption{ \label{fig:cit} Degree distribution of a citation network over different times}
\end{figure}

In order to reveal the limitations of the existing methods and the main ideas of our proposed method, Figure \ref{fig:cit_col} and Figure \ref{fig:cit} provide intuitive examples of networks with different sizes over time. Figure \ref{fig:cit_col} illustrates the degree distribution of two citation networks and two collaboration (co-authorship) networks which are extracted from CiteSeerX digital library \cite{ref45}. The networks represent two snapshots in the years 1992 and 2010. For example, \textit{Citation\_1992} represents the graph of citations in the papers of the CiteSeerX repository which are published before 1992. The figure shows two similar distributions for the two citation networks (resembling a power-law distribution) and two similar distributions for the two collaboration networks (similar to log-normal model). Obviously, the degree distributions of the citation networks are dissimilar to those of the collaboration networks. The existing similarity functions are unable to capture such similarity and dissimilarity in the shape of the degree distributions appropriately. For example, the Kolmogorov-Smirnov test will return the least similarity between the two citation networks, the power-law exponents are uninformatively different for the four networks, and the percentile quantification method \cite{ref20} returns an equal vector for all the four networks. Although the shape of the degree distributions in this example clearly reveals the similarity of the networks, in many situations the similarity of the degree distributions are not that obvious. Therefore, we need an automatic method for feature extraction and/or quantified network comparison using the degree distributions. It is worth noting that many real networks do not follow a specific distribution model such as the power-law or log-normal models. As a result, fitting the degree distribution to some predefined distribution models is not a good approach for feature extraction or comparison of the networks. 

As another example, Figure \ref{fig:cit} shows the degree distribution of the citation networks, extracted from the same CiteSeerX repository for different snapshots from 1990 to 2010. These networks have become bigger over years, with 191,443 nodes (papers) in 1990 and 1,039,952 nodes in 2010. Although all the networks show similar degree distributions, the bigger networks contain a wider range of node degrees. As the network size increases, maximum node degree, average degree, and standard deviation of the node degrees also increase. As a result, it seems that a degree distribution should be normalized according to its mean and deviation so that the comparison of the networks with different scales become valid. We follow this idea in our proposed methods of feature extraction and comparison of the network degree distributions. In our proposed ``feature extraction'' method, a fixed-length feature vector is extracted from the degree distribution, which can be used in data analysis applications, data-mining algorithms and comparison of degree distributions. In the ``comparison method'', we propose a distance function that computes the distance (amount of dissimilarity) between two given network degree distributions. With such a distance function, we can figure out how similar the given networks are, according to their degree distributions. 

Although our proposed approach is of general nature and equally applicable to other network types, in this paper we focus on simple undirected networks. In the rest of this paper, our proposed method is called ``Degree Distribution Quantification and Comparison (DDQC)''. The rest of this paper is organized as follows: In section \ref{Related Works}, we briefly overview the related works. In section \ref{Proposed Method}, we propose a new method for degree distribution quantification and comparison. In section \ref{Evaluation}, we evaluate the proposed method and we compare it with baseline methods. Finally, we conclude the paper in section \ref{Conclusion}.

\section{Related Works}\label{Related Works}
The degree distribution of many real-world networks are heavy tailed \cite{ref8, ref9, ref10, ref11, ref12}, and the power-law distribution is the most suggested model for complex networks\cite{ref4, ref8, ref10}. In power-law degree distribution the number of nodes with degree $d$ is proportional to $d^{-\gamma}$ ($ N_{d} \varpropto d^{-\gamma} $) where $\gamma$ is a positive number called ``the power-law exponent''. The value of $\gamma$ is typically in the range $2 < \gamma < 3$ \cite{ref2,ref37,ref38}. The fitted power-law exponent can be used to characterize graphs \cite{ref26}. A common approach for feature extraction from the degree distribution is to fit it on a power-law model and estimate the power-law exponent ($\gamma$). As a result, it will be possible to compare networks according to their fitted power-law exponents. One of the drawbacks of this approach is that power-law exponent is too limited to represent a whole degree distribution. This approach also follows the assumption that the degree distribution is power-law, which is not always valid, because many networks follow other degree distribution models  such as  log-normal distribution  \cite{ref13,ref14,ref38,ref39}. In addition, the power-law exponent does not reflect the deviation of the degree distribution from the fitted power-law distribution. As a result, two completely different distributions may have similar quantified feature (fitted power-law exponent).

An alternative approach for comparing the degree distributions is to utilize statistical methods of probability distribution comparison. Degree distribution is a kind of probability distribution and there are a variety of measures for calculating the distance between two probability distributions. In this context, the most common method is the Kolmogorov-Smirnov (KS) test, which is defined as the maximum distance between the cumulative distribution functions (CDF) of the two probability distributions \cite{ref8}. KS-test is used for comparing two degree distributions (two-sample KS test) \cite{ref11,ref23} and also for comparing a degree distribution with a baseline (usually the power-law) distribution \cite{ref8,ref13,ref40,ref41}. The KS distance of two distributions is calculated according to Equation \ref{eq:KS}, in which $S_{1}(d)$ and  $S_{2}(d)$ are the CDFs of the two degree distributions, and $d$ indicates the node degree. KS-test is largely utilized in the literature for comparing degree distribution of complex networks \cite{ref8,ref10,ref11,ref23,ref24,ref25}. KS-test is a method for comparing the degree distributions and calculating their distance, and it does not provide feature extraction mechanism. As a result, we should maintain the CDF of the degree distributions so that we can compare them according to KS-test. This is a drawback of KS-test since in other investigated approaches the degree distribution is summarized in a small fixed-length feature vector. Additionally, the KS-test is not applicable in data analysis applications that rely on feature vector representation of networks. KS-test is also sensitive to the scale and size of the networks, since it performs a point-to-point comparison of CDFs. Therefore, for two networks with different ranges of node degrees, the KS-test may return a large value as their distance even if the overall views of the degree distributions are similar (refer to Figure \ref{fig:cit_col}).

\small
\begin{equation} \label{eq:KS}
distance_{KS}(S_{1},S_{2})=\max_{d}\vert S_{1}(d)-S_{2}(d) \vert
\end{equation}
\normalsize

Janssen et. al., \cite{ref20} propose an alternative method for feature extraction from  the degree distribution. In this method, the degree distribution is divided into eight equal-size regions and the sum of degree probabilities in each region is extracted as distribution percentiles. This method is sensitive to the range of node degrees and also to outlier values of degrees. We recall this technique as ``Percentiles'' and we include it in baseline methods, along with ``KS-test'' and ``Power-law'' (the power-law exponent) in order to evaluate our proposed distance metric which is called ``DDQC''.

\section{Proposed Method}
\label{Proposed Method}

The degree distribution of a network is described in Equation \ref{eq:PG} as a probability distribution function. The equation shows the probability of node degrees in the given graph $G$, in which $D(v)$ is the degree of node $v$, and $P_G (d)$ is the probability that the degree of a node is equal to $d$. Based on our observations in networks of different sizes, we propose a new method for feature extraction and comparison of the degree distributions. In this method, a vector of eight real numbers is extracted from the degree distribution. A distance function is also suggested for comparing the feature vectors. In order to reduce the impact of the network size on the extracted features, we considered the mean and standard deviation of the degree distribution in the feature extraction procedure. Equation \ref{eq:meanG} and  Equation \ref{eq:stdxG} show the mean and standard deviation of the degree distribution respectively.

According to Equation \ref{eq:regions}, for any network $G$, we divide the range of node degrees into four regions ($R_G (r) , r=1..4$). The regions are defined based on the mean, standard deviation, minimum, and maximum of node degrees. Equation \ref{eq:regionlength} shows the length of each region ($|R_G (r)|$), in which $left(R)$ indicates the lower-bound (minimum degree) in region $R$ and $right(R)$ is its upper-bound. As Equation \ref{eq:intervals} shows, each region is further divided into two equal-size intervals ($I_G (i) , i=1..8$). Although it is possible to consider more than two intervals in each region, our experiments showed that considering more than two intervals per region results in no considerable improvements in the accuracy of the distance metric. Equation \ref{eq:idp} defines the ``interval degree probability'' ($IDP_G$) which shows the probability that the specified interval $I$ includes the degree of a randomly chosen node. We have defined eight intervals in the degree distribution (four regions and two intervals per region), and the degree distribution can be quantified based on the $IDP$ of the eight intervals. Equation \ref{eq:quantification} shows the final quantification (feature-vector) of the degree distribution. The proposed feature extraction method can be utilized in network analysis applications which rely on feature extraction and/or comparison of network degree distributions.

\small
\begin{equation} \label{eq:PG}
P_G (d)= P(D(v)=d );v\in V(G)
\end{equation}

\begin{equation} \label{eq:meanG}
\mu _G =\sum\limits_{d=min_G (D(v))}^{max_G (D(v))} d \times P_G (d)
\end{equation}

\begin{equation} \label{eq:stdxG}
\sigma _G =\sqrt{\sum\limits_{d=min_G (D(v))}^{max_G (D(v))}  P_G (d)\times(d-\mu _G)^2}
\end{equation}

\begin{equation} \label{eq:regions}
R_G (r) = 
\begin{cases} 
\big[ min_G (D(v)), \mu_G - \sigma_G \big]  & r=1 \\
\big[ \mu_G -   \sigma_G , \mu_G \big]  & r=2 \\
\big[ \mu_G , \mu_G +   \sigma_G \big] & r=3 \\
\big[ \mu_G +   \sigma_G  , max_G (D(v)) \big] & r=4.
\end{cases}
\end{equation}

\begin{equation}  \label{eq:regionlength}
\vert R_G (r)\vert = ⁡max \Big(right \big( R_G (r) \big) - left \big( R_G (r) \big),0 \Big	)
\end{equation}

\begin{equation} \label{eq:intervals}
I_{G}(i)=\\
\left\{
	\begin{matrix}
		\big[ left(R_{G}(\lceil\frac{i}{2}\rceil)), left(R_{G}(\lceil\frac{i}{2}\rceil))+\frac{\vert R_{G}(\lceil\frac{i}{2}\rceil) \vert }{2} \big] &\mbox{i is odd}
		\\
		\big[ left(R_{G}(\lceil\frac{i}{2}\rceil))+\frac{\vert R_{G}(\lceil\frac{i}{2}\rceil) \vert }{2} , right(R_{G}(\lceil\frac{i}{2}\rceil)) \big] &\mbox{i is even}
	
	\end{matrix}\right.
\end{equation}

\begin{equation} \label{eq:idp}
IDP_G (I)=P(left(I)\leq D(v) < right(I));   v\in V(G)
\end{equation}

\begin{equation} \label{eq:quantification}
Q(G)=\left\langle IDP_G (I_G (i)) \right\rangle  _{i=1..8}
\end{equation}

\normalsize

After the feature extraction phase, we can compare the degree distribution of two networks $G_1$ and $G_2$ according to their quantified feature vectors. We propose the Equation \ref{eq:distance} for comparing two degree distributions. This equation compares two networks based on their corresponding feature vectors ($Q_i$ values). Intuitively, $d(G_1 , G_2 )$ compares the corresponding interval degree probabilities of the two networks and sums their differences. Equation \ref{eq:distance} is a distance function for degree distribution of networks, and it is the result of a comprehensive study of different real, artificial and temporal networks.

\small
\begin{equation} \label{eq:distance}
d(G_1,G_2)=distance(G_1,G_2)=\\
	\sum\limits_{i=1}^{8} 
	\vert Q_i (G_1) - Q_i (G_2)  \vert
\end{equation}
\normalsize

\section{Evaluation}
\label{Evaluation}
In this section, we evaluate our proposed method. In subsection \ref{Datasets} we describe different network datasets which are used in our evaluations and in subsection \ref{Evaluation Results} we compare our proposed method with baseline methods.

\subsection{Datasets}
\label{Datasets}
In our problem setting, we aim a distance function that given the degree distribution of two networks, calculates how similar they are. But what does this ``similarity'' mean for degree distributions? What benchmark is available for evaluating such a distance function? For evaluating different distance metrics, an approved dataset of networks with known distances of its instances is sufficient. Although there is no such an accepted benchmark of networks with known ``distance values'', there exist some similarity witnesses among the networks. For evaluating different distance metrics, we have prepared two network datasets with admissible similarity witnesses among the networks of these datasets:
\begin{itemize}
\setlength{\itemindent}{-1em}
\item 
\textbf{\textit{Artificial Networks}}. 
We generated a dataset of 6,000 labeled artificial networks using six different generative models. Generative models are network generation methods for synthesizing artificial graphs that resemble the topological properties of real-world complex networks. We considered six generative models in the evaluations: Barab\'{a}si-Albert model \cite{ref7}, Erd\H{o}s-R\'{e}nyi \cite{ref44}, Forest Fire \cite{ref9}, Kronecker model \cite{ref29}, random power-law \cite{ref30}, and Small-world (Watts-Strogatz) model \cite{ref5}. Each evaluation scenario consists of 100 iterations of network generation and the average of the evaluation results of the 100 iterations are reported. In each iteration 60 networks are generated using the six generative models (10 networks per model). As a result, the evaluations on the artificial networks includes the generation of a total of 6,000 network instances using different parameters. The number of nodes in generated networks ranges from 1,000 to 5,000 nodes with the average of 2,936.34 nodes in each network instance. The average number of edges is 13,714.75. In this dataset, the generative models (generation methods) are the witnesses of the similarity: The networks generated by the same model follow identical link formation rules, and their degree distributions are considered similar. The networks of this data-set are further described in the \ref{datasetsappendix}, along with an overview of the selected generative models.
\item
\textbf{\textit{Real-world Networks}}. 
We have collected a dataset of 33 real-world networks of different types. The networks are selected from six different network classes: Friendship networks, communication networks, collaboration networks, citation networks, peer to peer networks and graph of linked web pages. The category of networks is a sign of similarity: networks of the same type usually follow similar link formation procedures and produce similar degree distributions. So, when comparing two network instances, we expect the distance metric to return small distances (in average) for networks of the same type and relatively larger distances for networks with different types. The ``real-world networks'' data-set is described in the \ref{datasetsappendix}, along with the basic properties and the source of its networks.
\end{itemize}

We assume that the category of the networks in the artificial and real datasets is a sign of their similarity. Networks of the same type follow similar link formation procedures and produce networks with similar structures. Although it is possible for two different-class networks to be more similar than two same-class networks, we assume that the overall ``expected similarity'' among networks of the same class is more than the expected similarity of different-class networks. This definition of the network similarity which is based on the network types is frequently utilized in the literature \cite{ref61, ref20, ref21, portraitsEPL, NetSimileASONAM, MehlerSim, taxonomiesPRE2012}.

\subsection{Evaluation Results}
\label{Evaluation Results}
In the section \ref{Datasets}, we described our two network datasets and we introduced different signs and witnesses of similarities among networks of these datasets. We can consider these witnesses in the evaluation of the proposed method. In this subsection, we evaluate our proposed distance function and we compare it with baseline methods based on their consistency to mentioned witnesses of the similarity. 

\begin{figure}
	\centering
	\includegraphics[scale=0.3]{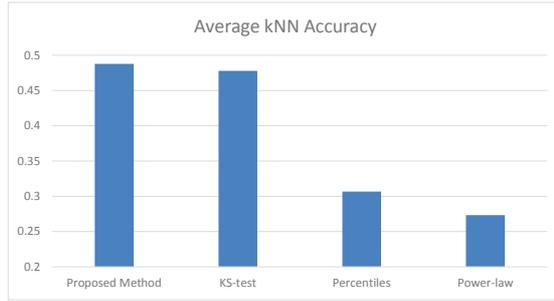}
	\caption{ \label{fig:knnreal} Average kNN accuracy in real networks dataset for K=1..10}
\end{figure}

\begin{figure}
	\centering
	\includegraphics[scale=0.3]{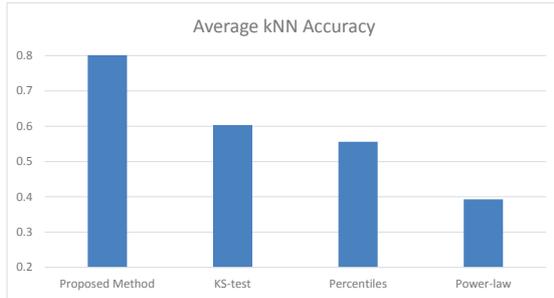}
	\caption{ \label{fig:knnart} Average kNN accuracy in artificial networks dataset for K=1..10}
\end{figure}

We start the evaluations by evaluating the precision of k-Nearest-Neighbor classifier based on different distance functions. The k-Nearest-Neighbor rule (kNN) \cite{ref46} is a common method for classification. It categorizes an unlabeled example by the majority label of its k-nearest neighbors in the training set. The performance of kNN is essentially dependent on the way that similarities are computed between different examples. Hence, better distance metrics result in better classification accuracy of kNN. In order to evaluate the accuracy of different distance functions, we employ them in kNN classification and we test the accuracy of this classifier. This evaluation is performed for both labeled datasets of real-world and artificial networks. As Equation \ref{eq:knnaccuracy} shows,  in a $dataset$ of labeled instances, the KNN-accuracy of a distance metric $d$ is the probability that the predicted class of an instance is equal to its actual class, when the distance metric $d$ is used in the KNN classifier. Figure \ref{fig:knnreal} and Figure \ref{fig:knnart} illustrate this evaluation in the real and artificial network datasets respectively. In these evaluations, kNN rule is performed with different values of $k$ from 1 to 10 and the average kNN accuracy is computed. As the figures show, the proposed method results in the best kNN accuracy among the baseline methods.

\small
\begin{equation}\label{eq:knnaccuracy}
KNN\textrm{-}Accuracy (d) = P( KNN\textrm{-}Classify_d (x) = class (x)) , x\in dataset
\end{equation}
\normalsize

\begin{figure}
	\centering
	\includegraphics[scale=0.4]{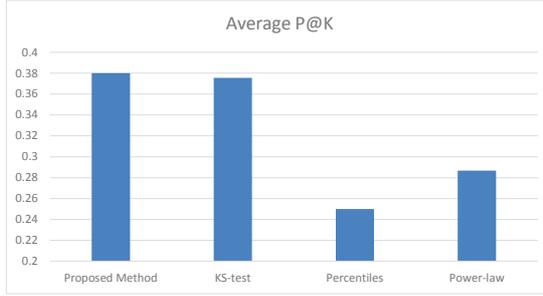}
	\caption{ \label{fig:pATk_real} Average P@K in real networks dataset for K=1..10}
\end{figure}

\begin{figure}
	\centering
	\includegraphics[scale=0.3]{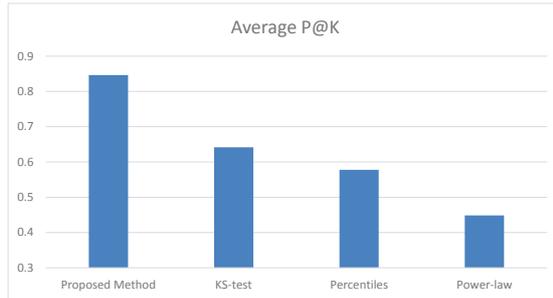}
	\caption{ \label{fig:pATk_art} Average P@K in artificial networks dataset for K=1..10}
\end{figure}

As an alternative evaluation criterion, we can consider the Precision-at-K (P@K) for assessing the proposed method. As Equation \ref{eq:pATk} shows, P@K indicates the percentage of classmates in the $K$ nearest neighbors. In other words, P@K is equal to the average number of classmates in the $K$ nearest neighbors of an instance, divided by $k$. P@K is dependent on the distance metric $d$ that is utilized for computing the distances among the dataset instances. Figure \ref{fig:pATk_real} and Figure \ref{fig:pATk_art} show the average P@K for K=1..10 in the networks of the real and artificial network datasets respectively, according to different distance metrics. As both the figures show, the proposed method outperforms all the baseline methods with respect to the average P@K measure.

\small
\begin{equation}\label{eq:pATk}
P@K(d)  = \frac{E(c)}{k} ; c=count (m), m \in KNN_d(x)  \; and \;  class(m)=class(x ) , x \in dataset 
\end{equation}
\normalsize

\begin{figure}
	\centering
	\includegraphics[scale=0.3]{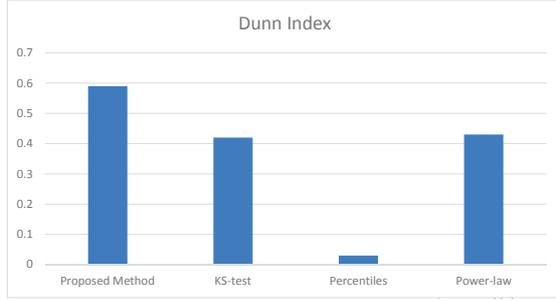}
	\caption{\label{fig:dunn_real}Dunn index for different distance metrics in real networks dataset.}
\end{figure}

\begin{figure}
	\centering
	\includegraphics[scale=0.3]{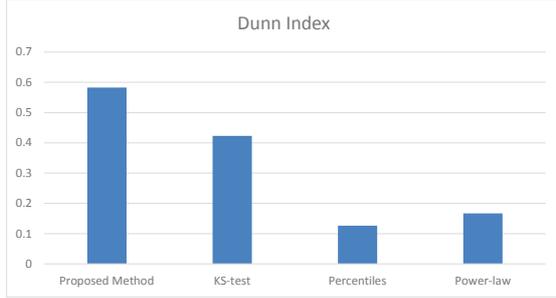}
	\caption{\label{fig:dunn_art}Dunn index for different distance metrics in artificial networks dataset.}
\end{figure}

An appropriate distance metric should return smaller distances for networks of the same class. In this context, Dunn Index \cite{GeneralizedDunn} is an appropriate measure for comparing the inter/intra class distances. For any partition setting $U$, in which the set of instances are clustered or classified into $c$ groups ($U \leftrightarrow X = X_1 \cup X_2   \cup ... \cup X_c $), Dunn defined the \textit{separation index} of $U$ as described in Equation \ref{eq:dunnIndex} \cite{GeneralizedDunn}. Dunn index investigates the ratio between the average distance of the two nearest classes (Equation \ref{eq:interDistance}) and the average distance between the members of the most extended class (Equation \ref{eq:intraDistance}). Figure \ref{fig:dunn_real} and Figure \ref{fig:dunn_art} illustrate the Dunn index for different network distance functions in the real and artificial network datasets respectively. The proposed method shows the best (the largest) Dunn index among the baseline methods.

\small
\begin{equation} \label{eq:dunnIndex}
DI(U) = \underset{1\leq i \leq c}{\underbrace{min}}\{ \underset{\underset{j \neq i}{1\leq j \leq c}}{\underbrace{min}} \{\frac{ \delta (X_{i},X_{j}) } {\underset{1 \leq k \leq c}{\underbrace{max}} \{\Delta (X_{k})\}} \} \}
\end{equation}

\begin{equation}\label{eq:interDistance}
\delta(S,T)=\delta_{avg}(S,T)=\frac{1}{|S||T|}\sum_{x \in S, y\in T}d(x,y)
\end{equation}

\begin{equation}\label{eq:intraDistance}
\Delta(S)=\frac{1}{|S|\cdot (|S|-1)}\sum_{x,y \in S , x \neq y}d(x,y)
\end{equation}

\normalsize

\begin{figure}
	\centering
	\includegraphics[scale=0.3]{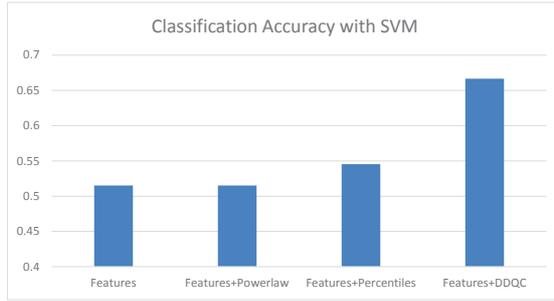}
	\caption{\label{fig:classification_real}SVM classification accuracy in real networks dataset.}
\end{figure}

\begin{figure}
	\centering
	\includegraphics[scale=0.3]{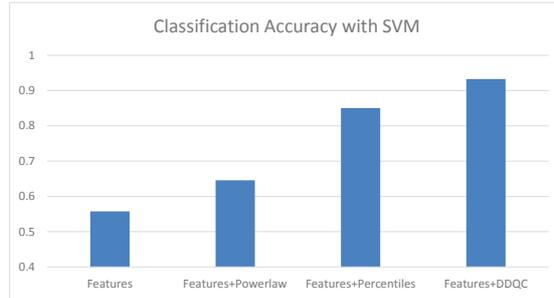}
	\caption{\label{fig:classification_art}SVM classification accuracy in artificial networks dataset.}
\end{figure}

In the last experiment, we investigate the integration of the degree distribution features with other network features, such as clustering coefficient and assortativity. In this experiment, we represent the networks with feature vectors that not only include the degree distribution features, but also some other important features that reflect the structural characteristics of complex networks. Then, we utilize supervised machine learning algorithms in order to classify the networks of the artificial and real network datasets using the integrated feature vectors. The aim of this experiment is to find the degree distribution features that best improves the accuracy of the classifier. Along with the degree distribution features, we consider four well-known network features: 1- ``Average clustering coefficient'' \cite{ref5} reflects the transitivity of connections. 2- ``Average path length'' \cite{ref3} shows the average shortest path between any pair of nodes. 3- The ``assortativity'' measure \cite{AssortativityNewman} shows the degree correlation between pairs of linked nodes . 4- ``Modularity'' \cite{ref6} is a measure for quantifying community structure of a network. When a network is represented with a feature vector of these four properties, we call the feature vector ``Features''. We consider three other feature vector representations in which the degree distribution features are also considered: ``Features+Powerlaw'' adds the fitted power-law exponent, ``Features+Percentiles'' includes the eight percentile features and ``Features+DDQC'' adds our proposed features of degree distribution. Since Kolmogorov-Smirnov (KS) does not provide a feature extraction method, we do not consider it in this experiment. We used Support Vector Machines (SVM) \cite{SMORef} as the classification method with 10-fold cross-validation. SVM performs a classification by mapping the inputs into a high-dimensional feature space and constructing hyperplanes to categorize the data instances. We utilized Sequential Minimal Optimization (SMO) \cite{SMORef} which is a common method for solving the optimization problem. Figure \ref{fig:classification_real} shows the accuracy of the classifier based on the described four versions of the feature vectors in the real networks dataset. Figure \ref{fig:classification_art} shows the result of the same experiment for the artificial networks dataset. As both the figures show, among the feature extraction methods for degree distribution, our proposed method results in the best classifier. In other words, our proposed method extracts the most informative features from the degree distribution that can improve the accuracy of network classification.

\section{Conclusion}
\label{Conclusion}

In this paper, we proposed a novel method for quantification (feature extraction) and comparison of network degree distributions. The aim of ``quantification'' is extracting a fixed-length feature vector from the degree distribution. Such a feature vector is used in network analysis applications such as network comparison. The ``network comparison'' is performed by returning a real number as the distance between two degree distributions. The distance is the counterpart of ``similarity'' and larger distances indicate less similarity. The degree distribution is an indicator of the link formation process  in the network, which reflects the overall pattern of connections \cite{ref27}. Similarly evolving networks have analogous degree distributions and we derive the similarity of degree distributions according to the similarity of link formation process in the networks. For deriving the amount of similarity of networks, we introduced admissible witnesses for network similarity: Similarity among the networks in two categories (same-type real networks and same-model artificial networks). We assume the networks in each of these categories have similar degree distributions. This assumption is the base of our evaluations for different degree-distribution distance metrics. Our proposed method, named DDQC, outperforms the baseline methods with regard to its accuracy in various evaluation criteria. The evaluations are performed based on different criteria such as the ability of the similarity metric to classify networks and comparison of inter/intra class distances. Although the integration of other network features (such as assortativity and modularity) improves the accuracy of the network similarity function, a similarity metric which is only based on the degree distribution has independent and important applications.

Our proposed method enables the data analysis applications and data mining algorithms to employ the degree distribution as a fixed-length feature vector. Hence, it is now possible to represent a network instance with a record of features (including clustering coefficient, average path length and the quantified degree distribution) and use such records in data analysis applications. As the future works, we will combine different network features along with the quantified degree distribution in an integrated distance metric for complex networks. Such an integrated distance metric will be the main building block of our future researches in evaluation and selection of network generative models and sampling methods. 

\section*{Acknowledgements}
We appreciate Sadegh Motallebi, Sina Rashidian and Javad Gharechamani for their helps in implementation and evaluation of the methods and in preparation of network datasets. We also thank Masoud Asadpour, Mahdieh Soleymani Baghshah and Hossein Rahmani for their valuable comments.

\appendix

\section{Overview of Artificial and Real-world Network Datasets }
\label{datasetsappendix}
In this appendix, we briefly describe the utilized network datasets of this research. The ``artificial networks'' dataset consists of a total of 6,000 networks, which are synthesized using six generative models. For each generative model, 1000 network instances are generated using different parameters. The number of nodes in generated networks is configured from 1,000 to 5,000 nodes, with the average of 2,936.34 nodes and 13,714.75 edges in each network instance. The selected generative models are some of the important and widely used network generation methods which cover a wide range of degree distribution structures. The generative models are described in the following, along with their configuration parameters in generation of ``artificial networks'' dataset. The values of the model parameters are selected according to the hints and recommendations of the cited original papers, along with a concern of keeping the number of network edges balanced for different models. The datasets are available upon request.

\begin{itemize}
\setlength{\itemindent}{-1em}
\item 
\textbf{\textit{Barab\'{a}si-Albert model (BA)}}. 
This is the classical preferential attachment model which generates scale free networks with power-law degree distributions \cite{ref7}. In this model, new nodes are incrementally added to the graph, one at a time. Each new node is randomly connected to $k$ existing nodes with a probability that is proportional to the degree of the available nodes. In the artificial networks dataset, $k$ is randomly selected as an integer number from the range $1\leqslant k \leqslant 10$.
\item 
\textbf{\textit{Erd\H{o}s-R\'{e}nyi (ER)}}. 
This model generates completely random graphs with a specified density \cite{ref44}. Network density is defined as the ratio of the existing edges to potential edges. The density of the ER networks in the artificial networks dataset is randomly selected from the range $ 0.002 \leqslant density \leqslant 0.005$.
\item 
\textbf{\textit{Forest Fire (FF)}}. 
This model, in which edge creation is similar to fire-spreading process, supports shrinking diameter and densification properties along with heavy-tailed in-degrees and community structure \cite{ref9}. This model is configured by two main parameters: Forward burning probability ($p$) and backward burning probability ($p_b$). For generating artificial networks dataset, we fixed $p_b=0.32$ and selected $p$ randomly from the range $0 \leqslant p \leqslant 0.3$.
\item 
\textbf{\textit{Kronecker graphs (KG)}}. 
This model generates realistic synthetic networks by applying a matrix operation (the kronecker product) on a small initiator matrix \cite{ref29}. The model is mathematically tractable and supports many network features including small path lengths, heavy tail degree distribution, heavy tails for eigenvalues and eigenvectors, densification, and shrinking diameters over time. The KG networks of the artificial networks dataset are generated using a $2 \times 2$ initiator matrix. The four elements of the initiator matrix are randomly selected from the ranges: $0.7 \leqslant P_{1,1} \leqslant 0.9, 0.5 \leqslant P_{1,2} \leqslant 0.7, 0.4 \leqslant P_{2,1} \leqslant 0.6, 0.2 \leqslant P_{2,2} \leqslant 0.4$.
\item 
\textbf{\textit{Random power-law (RP)}}. 
This model follows a variation of ER model and  generates synthetic networks with power law degree distribution \cite{ref30}. This model is configured by the power-law degree exponent ($\gamma$). In our parameter setting for generating artificial networks dataset, $\gamma$ is randomly selected from the range $2.5 < \gamma <3$.
\item 
\textbf{\textit{Watts-Strogatz model (WS)}}. 
The classical Watts-Strogatz small-world model synthesizes networks with small path lengths and high clustering \cite{ref5}. It starts with a regular lattice, in which each node is connected to $k$ neighbors, and then randomly rewires some edges of the network with rewiring probability $\beta$. In WS networks of the artificial networks dataset, $\beta$ is fixed as $\beta =0.5$, and $k$ is randomly selected from the integer numbers between 2 and 10 ($2 \leqslant k \leqslant 10$).
\end{itemize}

Table \ref{tab:realnets} describes the graphs of the ``real-world networks'' dataset, along with the category, number of nodes and edges, and the source of these graphs. Most of these networks are publicly available datasets. Two temporal networks (Cit\_CiteSeerX and Collab\_CiteSeerX) are extracted from CiteSeerx digital library \cite{ref45}, using a web crawler software tool. 

\begingroup
\begin{table*}
	\centering
\caption{ \label{tab:realnets} Dataset of real-world networks}

    \begin{tabular}{|c|lrrr|}
    \hline
    \textbf{	Category} & \textbf{ID} & \textbf{Vertices} & \textbf{Edges} & \textbf{Source} \\
    \hline
    \multirow{4}{*}{Citation Network} & Cit-HepPh & 34,546 & 420,899 & SNAP \cite{ref50} \\
\cline{2-5}          & Cit-HepTh & 27,770 & 352,304 & SNAP \cite{ref50} \\
\cline{2-5}          & dblp\_cite & 475,886 & 2,284,694 & DBLP \cite{ref51} \\
\cline{2-5}          & Cit\_CiteSeerX & 1,106,431 & 11,791,228 & CiteSeerX \cite{ref45} \\
    \hline
    \multirow{10}{*}{Collaboration Network} & CA-AstroPh & 18,772 & 198,080 & SNAP  \cite{ref50} \\
\cline{2-5}          & CA-CondMat & 23,133 & 93,465 & SNAP  \cite{ref50} \\
\cline{2-5}          & CA-HepTh & 9,877 & 25,985 & SNAP  \cite{ref50} \\
\cline{2-5}          & Collab\_CiteSeerX & 1,260,292 & 5,313,101 & CiteSeerX \cite{ref45} \\
\cline{2-5}          & com-dblp & 317,080 & 1,049,866 & SNAP  \cite{ref50} \\
\cline{2-5}          & dblp\_collab & 975,044 & 3,489,572 & DBLP \cite{ref51} \\
\cline{2-5}          & dblp20080824 & 511,163 & 1,871,070 & Sommer \cite{ref52} \\
\cline{2-5}          & IMDB-09 & 4,155 & 16,679 & Rossetti \cite{ref53} \\
\cline{2-5}          & CA-GrQc & 5,242 & 14,490 & SNAP  \cite{ref50} \\
\cline{2-5}          & CA-HepPh & 12,008 & 118,505 & SNAP  \cite{ref50} \\
    \hline
    \multirow{4}{*}{Communication Network} & EmailURV & 1,133 & 5,451 & Aarenas \cite{ref54} \\
\cline{2-5}          & Email-Enron & 36,692 & 183,831 & SNAP \cite{ref50,ref55} \\
\cline{2-5}          & Email-EuAll & 265,214 & 365,025 & Konect \cite{ref56} \\
\cline{2-5}          & WikiTalk & 2,394,385 & 4,659,565 & SNAP \cite{ref50} \\
    \hline
    \multirow{7}{*}{Friendship Network} & Dolphins & 62    & 159   & NetData \cite{ref57} \\
\cline{2-5}          & facebook-links & 63,731 & 817,090 & MaxPlanck \cite{ref58} \\
\cline{2-5}          & Slashdot0811 & 77,360 & 507,833 & SNAP \cite{ref50} \\
\cline{2-5}          & Slashdot0902 & 82,168 & 543,381 & SNAP \cite{ref50} \\
\cline{2-5}          & soc-Epinions1 & 75,879 & 405,740 & SNAP \cite{ref50} \\
\cline{2-5}          & Twitter-Richmond & 2,566 & 8,593 & Rossetti \cite{ref53} \\
\cline{2-5}          & youtube-d-growth & 1,138,499 & 2,990,443 & MaxPlanck \cite{ref58} \\
    \hline
    \multirow{4}{*}{Graph of Web Pages} & web-BerkStan & 685,230 & 6,649,470 & SNAP \cite{ref50} \\
\cline{2-5}          & web-Google & 875,713 & 4,322,051 & SNAP \cite{ref50} \\
\cline{2-5}          & web-NotreDame & 325,729 & 1,103,835 & SNAP \cite{ref50} \\
\cline{2-5}          & web-Stanford & 281,903 & 1,992,636 & SNAP \cite{ref50} \\
    \hline
    \multirow{4}{*}{P2P Network} & p2p-Gnutella04 & 10,876 & 39,994 & SNAP \cite{ref50} \\
\cline{2-5}          & p2p-Gnutella05 & 8,846 & 31,839 & SNAP \cite{ref50} \\
\cline{2-5}          & p2p-Gnutella06 & 8,717 & 31,525 & SNAP \cite{ref50} \\
\cline{2-5}          & p2p-Gnutella08 & 6,301 & 20,777 & SNAP \cite{ref50} \\
    \hline
    \end{tabular}%
\end{table*}
\endgroup

\section*{References}

\bibliographystyle{elsarticle-num}
\bibliography{shortbib}

\end{document}